\UseRawInputEncoding
\documentclass[12pt, draftclsnofoot, onecolumn]{IEEEtran}
\normalsize
\usepackage{pifont,bm,multicol,amsfonts,amsmath,color,amssymb,graphicx, epsfig,cite,psfrag,subfigure,algorithm,enumerate,stfloats,algorithm,algorithmic,epstopdf,balance,multirow,multicol}

\newtheorem{remark}{\underline{Remark}}

\begin{document}
\title{Ordinary Differential Equation-based CNN for Channel Extrapolation over RIS-assisted Communication}

\author{Meng Xu, Shun Zhang, \emph{Senior Member, IEEE}, Caijun Zhong, \emph{Senior Member, IEEE},  Jianpeng Ma, \emph{Member, IEEE}, Octavia A. Dobre, \emph{Fellow, IEEE}

%
\thanks{M. Xu, S. Zhang and J. Ma are with the State Key Laboratory of Integrated Services Networks, Xidian University, Xi¡¯an 710071, P. R. China (e-mail: mxu$\_$20@stu.xidian.edu.cn, zhangshunsdu@xidian.edu.cn, jpmaxdu@gmail.com).}

\thanks{C. Zhong is with the College of Information Science and Electronic Engineering, Zhejiang University, Hangzhou  310027, P. R. China (e-mail: caijunzhong@zju.edu.cn).}



\thanks{O. A. Dobre is with Faculty of Engineering and Applied Science, Memorial University, St. John's NL AIC-5S7, Canada (e-mail: odobre@mun.ca).}
}

\maketitle
\vspace{-10mm}
\begin{abstract}
The reconfigurable intelligent surface (RIS) is considered as a promising new technology for reconfiguring wireless communication environments.
To acquire the channel information accurately and efficiently,
we only turn on  a fraction of all the RIS elements, formulate a sub-sampled RIS channel, and design a deep learning based scheme
to extrapolate the full channel information from the partial one.
Specifically, inspired by the ordinary differential equation (ODE), we
set up connections between different data  layers in a
convolutional neural network (CNN) and improve its structure.
Simulation results are provided to demonstrate that our proposed ODE-based CNN structure can achieve faster convergence speed and better solution than the cascaded CNN.
\end{abstract}

\maketitle
\thispagestyle{empty}
\vspace{-1mm}

\begin{IEEEkeywords}
Convolutional neural network, RIS, channel extrapolation, ordinary differential equation, sub-sample.
\end{IEEEkeywords}

\section{Introduction}
With the increasing demands for communication services, the number of connected devices continues to augment exponentially and new, blossoming service requirements pose more constraints on the network.
At the same time,  the increased power consumption and hardware cost remain key issues \cite{efficiency2}.
A recent technological breakthrough that holds the potential
to overcome these technological bottlenecks is
reconfigurable intelligent surface (RIS).
It can be coated on any environmental object in a cost-effective manner, thereby facilitating the large scale deployment. Due to the physical characteristics of RIS, it can reflect incident electromagnetic waves, and adjust their amplitude and phase in a controlled manner.
In other words, RIS can manipulate the communication
environment in an intelligent way.
Furthermore, the reflection elements of RIS usually work in a passive state, which makes RIS~have low power consumption~\cite{passive1}. Hence, RIS is considered as a promising technology and attracts more and more attention.

{Similar to the case of other communication systems \cite{CE_Ma}, the acquisition of channel
state information (CSI) is an important problem and has become a hot research topic in RIS-assisted communication systems.
In \cite{two_stage_estimation1}, Ardah \emph{et al.} designed a two-stage channel estimation framework with high resolution.
In \cite{two_stage_estimation2},
the atomic norm minimization was resorted to implement the channel
estimation over RIS-aided MIMO system in the millimeter wave frequency band.}

{All of the above works depend on the hypothetical statistical model. However, in the actual communication scenario, the radio scattering conditions
change rapidly with time and are very complicated. This makes the traditional methods have some limitations  \cite{CE_Ma}. With the development of the artificial intelligence, the application of deep learning
(DL) in RIS-aided systems has attracted extensive attention.
In \cite{deeplearning1}, the authors adopted fully connected neural networks
 to  estimate the RIS channel and detect the symbols.
In \cite{deeplearning3}, Elbir \emph{et al.} designed a twin convolutional neural network (CNN) to estimate the direct and the cascaded channel in a RIS-aided communication system.
However, due to the passive characteristics of RIS, the channels
from the source to RIS and that from RIS to destination are coupled, and
the size of the equivalent channel is in scale with the number of the RIS elements, which is
usually large enough to  accurately manipulate an
incoming electromagnetic (EM) field. Thus, it would cost many pilot resource to directly achieve
the equivalent channel of large size at destination. Thus, how to reduce the overhead of the channel estimation is an interesting topic over the RIS-aided network. Recently, Taha et {\it al.} used a small part of RIS elements  to sub-sample the channels, and optimize the beamforming vector of RIS with
the channel estimated at the selected elements~\cite{Taha}.

In this paper, we further examine the  channel compression over the physical space for RIS-aided communication. After selecting a fraction of the RIS elements, we achieve
the equivalent channel formed by the source, the destination and the chosen RIS elements. Then, we extrapolate the channels to all elements from those estimated at chosen elements, where DL is adopted.
Furthermore, inspired by the ordinary differential equation (ODE), we modify the structure of the cascaded CNN by adding cross-layer connections, namely introducing coefficients and linear calculations between the network layers.
The proposed ODE-based CNN can obtain more accurate solutions, and its performance can be verified to be better than the cascaded CNN.

\vspace{-2mm}
\section{System And Channel Model}
As shown in Fig. \ref{system scene}, let us consider an indoor scenario,
where multiple-antenna base station (BS) communicates with a single-antenna user via RIS reflection.
A BS is equipped with a uniform linear array (ULA) with $M$ antennas.
RIS is in the form of a uniform planar array (UPA) and consists of $L=L_hL_v$ elements, where $L_h$ and $L_v$ separately denotes the sizes along the horizontal and vertical dimensions.
Moreover, the orthogonal frequency division multiplexing (OFDM) scheme is adopted and the number of subcarriers is $K$.
{Since the indoor environment is easy to be blocked by objects or people, the direct channel between the BS and the user may be destroyed. Thus, we only consider the channel reflected by RIS, i.e., the cascaded channel, instead of the direct channel.}
Obviously, the cascaded channel consists of two parts: the link from the BS to the RIS, i.e., $\mathbf{H}\in\mathbb{C}^{L\times M}$, and that from the RIS to the user, i.e., $\mathbf{g^\mathrm{H}}\in\mathbb{C}^{L\times 1}$. The received signal at the $k$-th subcarrier of the  user can be given as
\begin{align}
y_k = \mathbf{g}_k\mathbf{\Psi}\mathbf{H}_k\mathbf{s}_k + n_k,
\end{align}
where $\mathbf{\Psi}\in\mathbb{C}^{L\times L}$ is a diagonal matrix, i.e., $\mathbf{\Psi} = \text{diag}\{\beta_1\text{exp}(j\phi_1),\dots,\beta_L\text{exp}(j\phi_L)\}$,
$\mathbf{s}_k$ is the downlink $M\times1$ transmitted signal at the $k$-th sub-carrier, and $n_k \sim \mathcal{CN}(0,\sigma_n^2)$ is the addictive white Gaussian noise.  Due to the lack of signal processing capability at RIS, $\boldsymbol{\Psi}$  is the same at different sub-carriers.
Notice that $\phi_i$ in $\mathbf{\Psi}$ represents the phase shift
introduced by each RIS element while $\beta_i$ controls this element's on-off state, which will be described
in the following. Moreover, the channel between BS and RIS at the $k$-th subcarrier is given by
\begin{align}
\mathbf{H}_k = \frac{1}{\sqrt{K}}\sum_{i=1}^{P_h}h_{i,f_c}e^{-j2\pi\frac{k\tau_{h,i}}{KT_s}}\mathbf{a}_r(\phi_{h,i},\theta_{h,i})
\mathbf{a}_t^{\mathrm{H}}(\psi_{h,i}),
\end{align}
where $h_{i,f_c}$ is the complex channel gain along the $i$-th scattering
path at the
carrier frequency $f_c$, $\tau_{h,i}$ is the time delay, and
$\mathbf{a}_r(\phi_{h,i},\theta_{h,i})$, $\mathbf{a}_t(\psi_{h,i})$ are the spatial steering vectors, with $\phi_{h,i}$ and $\theta_{h,i}$ as the azimuth angle and elevation angle of the receiver, respectively, and $\psi_{h,i}$ as the angle of departure (AoD).
Correspondingly, $\mathbf{a}_r(\phi_{h,i},\theta_{h,i})$ can be written as
\begin{align}
\mathbf{a}_r(\phi_{h,i},\theta_{h,i}) = \mathbf{a}_{el}(\phi_{h,i})\otimes \mathbf{a}_{az}(\phi_{h,i},\theta_{h,i})\in \mathcal{C}^{L\times 1},
\end{align}
where the $L_v\times 1$ vector $\mathbf{a}_{el}(\phi_{h,i}) = [1,e^{-j2\pi\frac{d}{\lambda_c}\cos\phi_{h,i}},\dots,e^{-j2\pi\frac{d}{\lambda_c}(L_v-1)\cos\phi_{h,i}}]^\mathrm{T}$ and the $L_h\times 1$ vector $\mathbf{a}_{az}(\phi_{h,i},\theta_{h,i}) = [1,e^{-j2\pi\frac{d}{\lambda_c}\sin\phi_{h,i}\cos\theta_{h,i}},\dots,e^{-j2\pi\frac{d}{\lambda_c}(L_h-1)
\sin\phi_{h,i}\cos\theta_{h,i}}]^\mathrm{T}$. $\lambda_c$ is the carrier wavelength and $d$ denotes antenna spacing. Furthermore, $\otimes$ represents the Kronecker product operator and $[\cdot]^\mathrm{T}$ represents the transpose.
Moreover, $\mathbf{a}_t(\psi_{h,i})$ can be~given~by
\begin{small}\begin{align}
\mathbf{a}_t(\psi_{h,i}) = \frac{1}{\sqrt{M}}[1,e^{j\frac{2\pi}{\lambda_c}d\sin\psi_{h,i}},\dots,e^{j\frac{2\pi}{\lambda_c}d(M-1)\sin\psi_{h,i}}]^\mathrm{T}.
\end{align}\end{small}
{Correspondingly, the channel between the RIS and the user at the $k$-th subcarrier is
\begin{align}
\mathbf{g}_k^\mathrm{H} = \frac{1}{\sqrt{K}}\sum_{i=1}^{P_g}g_{i,f_c}e^{-j2\pi\frac{k\tau_{g,i}}{KT_s}}\mathbf{a}_t^{\mathrm{H}}
(\phi_{g,i},\theta_{g,i}),
\end{align}
where the structure of $\mathbf{a}_t(\phi_{g,i},\theta_{g,i})$ is similar to that of $\mathbf{a}_r(\phi_{h,i},\theta_{h,i})$.}

The cascaded channel matrix between BS and the user at the $k$-th subcarrier can be defined as $\mathbf{C}_{k} = \mathbf{G}_k\mathbf{H}_k$, where $\mathbf{G}_k = \mathrm{diag}\{\mathbf{g}_k\}$, and $\mathbf{C}_{k}$ has a size of $L\times M$.
Then, let us define the $LM\times 1$ vector $\mathbf{c}_k = [\mathbf{c}_{k,1}^\mathrm{T},\dots,\mathbf{c}_{k,L}^\mathrm{T}]^{\mathrm{T}}$, where $\mathbf{c}_{k,i}$ represents the $i$-th column of $\mathbf{C}_k$.
Within the RIS communication system,  it is proved that the optimal $\boldsymbol\Psi$ is closely related with all the cascaded channels at $K$ subcarriers, i.e., $\mathbf{C} = [\mathbf{c}_1, \mathbf{c}_2, \dots, \mathbf{c}_K] $ \cite{beamforming}.
 Hence, our aim is to estimate~$\mathbf{C}$.

\begin{figure}[!t]
	\centering
	\includegraphics[width=100mm]{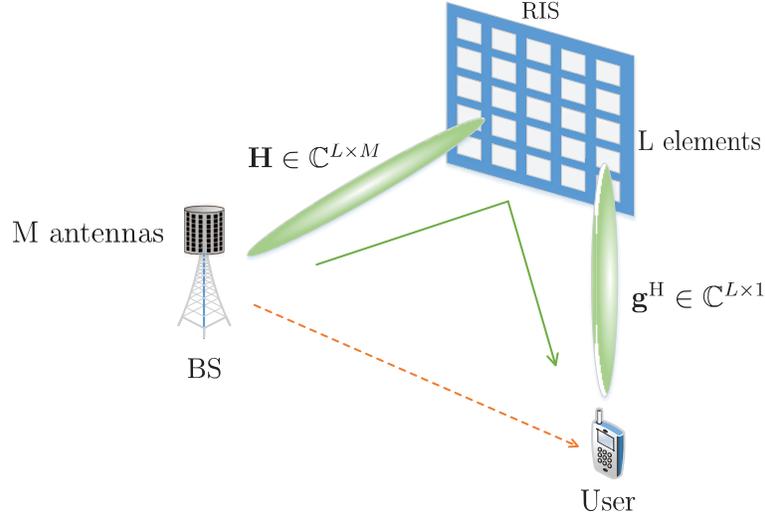}
	\caption{{RIS-aided communication system in an indoor scene.}}
	\label{system scene}
\end{figure}
\vspace{-2mm}

\section{Proposed Channel Extrapolation Method}

\subsection{Framework Design}
{

Theoretically, we can send a pilot matrix  $\mathbf{X}_k$ of size $ML\times S$ to directly recover $\mathbf c_k$ with the linear estimator, where
$S$ represents the time duration of $\mathbf X_k$. From the Bayesian estimation theory, %
 we can effectively recover $\mathbf c_k$ when  $S\geq ML$.
However, in massive MIMO systems, both $M$ and $L$ are relatively large.
Then, a significant number of pilot needs to be employed, which
drastically decreases the spectrum efficiency of the transmission.
To overcome this bottleneck, we can utilize a fraction of $N$ RIS elements and sub-sample $\mathbf C_k$.
Without loss of generality, we set the number of the selected RIS elements as $N$,
which can be implemented through  setting the parameters $\beta_i$, $\phi_i$ in $\boldsymbol\Psi$, $i=1,2,\ldots,L$.
{Specifically,
for the $N$  chosen elements, we set their $\phi_i$ and $\beta_i$ as $0$ and 1, respectively. For others, the corresponding amplitude
parameter $\beta_i$  is 0.
After this operation, the size of the sub-sampled cascaded channel at the $k$-th subcarrier is reduced to $N$.
Correspondingly, the sub-sampled cascaded channel at  $K$ subcarriers can be written as  $\widetilde{\mathbf{C}}\in\mathbb{C}^{MN\times K}$.
Obviously, compared with $\mathbf C$, a pilot sequence of shorter time duration would be required to estimate $\widetilde{\mathbf C}$.

If the power of the pilot sequence is large enough, we can estimate  $\widetilde{\mathbf C}$ with quite high accuracy.
However, we should utilize $\widetilde{\mathbf C}$ to infer the unknown cascaded channel at the $L-N$ non-chosen RIS elements. Thus, in
the following, we construct a DL-based framework  to extrapolate $\mathbf C$ from $\widetilde{\mathbf C}$. It may be worth noting that the selection pattern for the RIS elements
can impact the performance of the channel extrapolation and should be optimized. This topic is beyond the scope of this paper, though. Further, we adopt
the uniform sampling scheme.

%
\vspace{-2mm}
\subsection{ODE-based Channel Extrapolation}
As mentioned above, the input of the network is $\widetilde{\mathbf{C}}$, and its output is $\mathbf{C}$.
The task of our network is to learn the mapping function from $\widetilde{\mathbf{C}}$ to $\mathbf{C}$.
In other words, we want to estimate the complete channel through the sub-sample version, which is made possible by the correlation between different RIS elements.


The channel extrapolation  is similar to the super-resolution  in the field of image processing.
For this kind of problem, CNN has great advantages and is very suitable to use the correlation between data elements for information completion.
 In order to get better network performance, we can increase the number of data layers or modify the network structure.
 However, more layers will result in higher calculational requirements.
 Moreover, when the number of layers reaches a certain number, the improvement  become less and less.
 {Sometimes, the excessive deepening of the network causes the gradient explosion and disappearance. Thus,
 optimizing the network structure is more widely used than simply deepening the neural network.
 Theoretically, if we add some proper connections between layers, the performance of the network may be better, like residual neural network (ResNet) \cite{ResNet}.

Recently, ODE have been introduced to the neural network and utilized to describe
the latent relation between different data layers  \cite{NODE}.
With such powerful characterization, we could speed up the convergence and learning performance of the CNN.
Moreover, with the development of mathematical science, it is possible to use the numerical solutions of differential equations to modify the network structure and obtain possible gains.
Here, we incorporate two numerical approximation methods, i.e., LeapFrog and Runge-Kutta methods, into CNN.
The main difference between them lies in the approximation accuracy.

$\mathbf{LeapFrog \ method}$ :
LeapFrog method is a second-order approximation scheme and can be written as
\begin{align}
y_{n+1} = y_{n-1} + 2hf(x_n,y_n), \label{LeapFrog}
\end{align}
where $f(x_n,y_n)$ denotes the derivative at $(x_n,y_n)$ and $2h$ can be seen as an interval of width $x_{n+1} - x_{n-1}$.
Applying (\ref{LeapFrog}) for the CNN, we can connect the ($n+1$)-th layer with the ($n-1$)-th one. The corresponding relationship
can be formulated as:
\begin{align}
\mathbf{D}_n = \mathbf{D}_{n-2} + G(\mathbf{D}_{n-1}),\ n = 3,4,5,\dots,\label{LF}
\end{align}
where $\mathbf{D}_i$ represents the output data of the $i$-th layer, and $G(\cdot)$ is an operation containing a
 ReLu activation function, a convolution layer and a multiplier.

\begin{remark}The LeapFrog method is an improved version  of the forward Euler equation,
which can be written as $y_{n+1} = y_{n} + hf(x_n,y_n)$. The forward Euler method is the simplest first-order approximation of ODE
  and has a similar structure with ResNet.
\end{remark}

$\mathbf{Runge-Kutta \ methods}$ :
The theory of the numerical ODEs suggests that a higher-order approximation results in less truncation error and higher accuracy.
Hence, we turn to the Runge-Kutta methods, which are common numerical methods for ODE and can  be expressed as \cite{CVPR}
\begin{align}
y_{n+1} = y_{n} + \sum_{i=1}^{I}\gamma_iG_i, \label{RK}
\end{align}
and $G_1 = hf(x_n,y_n)$, while $G_i$ has the form of
\begin{align}
G_i = hf(x_n + \alpha_ih,y_n + \sum_{j=1}^{i-1}\beta_{ij}G_j),\ i = 2,3,\cdots,n,
\end{align}
where $I$ represents the number of stages;
$\alpha_i$ and $\beta_i$, like $\gamma_i$ in (\ref{RK}), are the related coefficients of the $i$-th stage.

If we set $\gamma_1=\frac{1}{6}, \gamma_2=\frac{2}{3}, \gamma_3=\frac{1}{6}, \alpha_2=\frac{1}{2}, \beta_{21}=\frac{1}{2}, \alpha_3=1, \beta_{31}=-1, \beta_{32}=2$, we obtain the 3-stage Runge-Kutta equation as the basic structure in our work, which can be expressed as
\begin{align}
y_{n+1} &= y_{n} + \frac{1}{6}(G_1 + 4G_2 + G_3), \label{RK3-formula-1}
\end{align}
where $G_1$, $G_2$, $G_3$ can be separately written as
\begin{align}
G_1 &= hf(x_n,y_n), \\
G_2 &= hf(x_n + \frac{h}{2},y_n + \frac{1}{2}G_1), \\
G_3 &= hf(x_n + h,y_n - G_1 + 2G_2). \label{RK3-formula-4}
\end{align}

\begin{figure}[!t]
	\centering
	\includegraphics[width=5in]{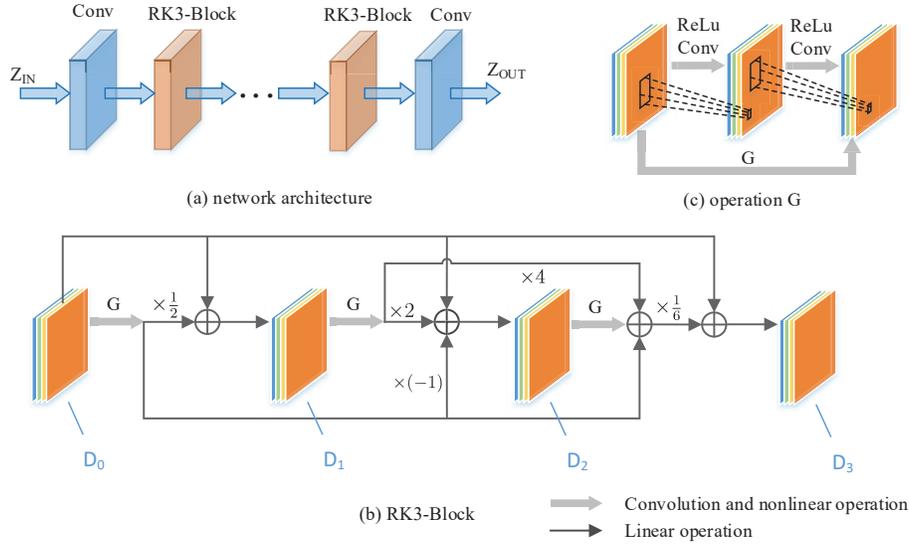}
	\caption{{(a) Our proposed network architecture; (b) The structure of the RK3-Block; (c) The architecture of operation $G$, which contains two ReLu activation functions and two convolutional layers.}}
	\label{system architecture}
\end{figure}

With (\ref{RK3-formula-1}) - (\ref{RK3-formula-4}),
we can construct an improved CNN structure, referred to as RK3-Block and depicted in Fig. \ref{system architecture} (b).
Correspondingly, the constraints among different layers in this block can be written~as
\begin{small}\begin{align}
\mathbf{D}_1 =& \  \mathbf{D}_0 + \frac{1}{2}G^{\prime}(\mathbf{D}_{0}),\
\mathbf{D}_2 = \mathbf{D}_0 - G^{\prime}(\mathbf{D}_{0}) + 2G^{\prime}(\mathbf{D}_{1}),\\
& \mathbf{D}_3 = \mathbf{D}_0 + \frac{1}{6}(G^{\prime}(\mathbf{D}_{0}) + 4G^{\prime}(\mathbf{D}_{1}) + G^{\prime}(\mathbf{D}_2)),
\end{align}\end{small}where the positions of $\mathbf D_i$, $i=0,1,2,3$, are presented in Fig. \ref{system architecture} (b), and the operation $G^{\prime}(\cdot)$ contains two ReLu activation functions and two convolution layers.
Similar to RK3-Block, with (\ref{LF}), we can obtain a modified CNN structure from the LeapFrog approximation and refer to it as LF-Block, which
 includes three data layers. As shown in Fig. \ref{system architecture} (a), we can cascade several RK3-Blocks or LF-Blocks to deepen the network
 for better results.

\vspace{-2mm}
\subsection{Learning Scheme}
The valid input of our network is the sub-sampled channel  $\widetilde{\mathbf{C}}\in\mathbb{C}^{MN\times K}$,
and the label is the entire cascaded channel $\mathbf{C}\in\mathbb{C}^{ML\times K}$.
In order to facilitate the training and the generation of data, we set the entries of $\mathbf C$ related with the $L-N$ non-chosen RIS elements
as 0 and obtain the resultant matrix $\mathbf C^{o}$, whose non-zero entries are  same with those of $\widetilde{\mathbf C}$.
Correspondingly, we treat $\mathbf C^{o}$ as the raw input of the ODE-based CNN.


Then, we reshape the raw input data and the label of the network as
$\mathbf{Z}_{\mathrm{IN}} = [\Re({\mathbf{C}^o});\Im({\mathbf{C}^o})]$ and $\mathbf{Z}_{\mathrm{TA}} = [\Re(\mathbf{C});\Im(\mathbf{C})]$,
respectively.
 Both $\mathbf{Z}_{\mathrm{IN}}$ and $\mathbf{Z}_{\mathrm{TA}}$ are real-valued matrices with the size of $ML\times K\times 2$. Correspondingly, the output of this network can be written as $\mathbf{Z}_{\mathrm{OUT}} = [\Re(\widehat{\mathbf{C}});\Im(\widehat{\mathbf{C}})] \in \mathbb{C}^{ML\times K\times 2}$,
where $\widehat{\mathbf{C}}$ is the estimate of ${\mathbf{C}}$.
{In our proposed network, there are $N_c$ convolutional layers. In the $n$-th layer, the input is processed by $N_k$ convolutional kernels of size $H\times W$. Note that $H$ and $W$ represent the height and the width of the convolutional kernels.
Normally, the size of the output data in each convolutional layer depends on $H$ and $W$, and it is usually slightly smaller than the input data.}

{During the learning stage, the  parameter vector
$\boldsymbol\omega = [\boldsymbol\omega_1^{T},\boldsymbol \omega_2^\mathrm{T},\dots,\boldsymbol\omega_{N_c}^{T}]^{T}$
is optimized by minimizing the mean squared error (MSE) between the output $\mathbf{Z}_{\mathrm{OUT}}$ and the target $\mathbf{Z}_{\mathrm{TA}}$,
where the vector $\boldsymbol\omega_{n}$ contains all the model parameters of the $n$-th layer, $n=1,2,\ldots,N_c$.
Hence, the loss function can be written as
\begin{align}
\mathcal{L} = \frac{1}{M_bMLK}\sum_{i=1}^{M_b}\begin{Vmatrix}[\mathbf{Z}_{\mathrm{TA}}]_i - [\mathbf{Z}_{\mathrm{OUT}}]_i \end{Vmatrix}_F^2,
\end{align}
where $||\mathbf{A}||_F$ is the $\mathrm{Frobenius}$ norm of matrix $\mathbf{A}$ and $M_b$ denotes the batch size for training.
Here, the adaptive moment estimation (Adam) \cite{ADAM} algorithm is adopted to achieve the best $\boldsymbol\omega$,
which is controlled by  the learning rate $\eta$.

\begin{table}[t]
		\centering
		\renewcommand{\arraystretch}{1.2}
		\caption{Layer Parameters for the CNN with ODE-RK3 Structure.}
		\label{layer parameters}
        \scalebox{1.2}{
		\begin{tabular}{c c c c c c  }
			\hline
			Layer &Output size &Activation &Kernel size &Strides \\
			\hline
            \hline
			$1\times$ Conv2D &$256\times 64\times 128$ &None &$5\times 5$ &$1\times 1$ \\
            \hline
			$4\times$ RK3-Block &$256\times 64\times 128$ &ReLu &$3\times 3$ &$1\times 1$ \\
            \hline
			$1\times$ Conv2D &$256\times 64\times 2$ &None &$3\times 3$ &$1\times 1$ \\
			\hline
            \hline	
		\end{tabular}}	
	\end{table}

\newcommand{\tabincell}[2]{\begin{tabular}{@{}#1@{}}#2\end{tabular}}
\renewcommand{\arraystretch}{1}
\begin{table}[t]
  \centering
  \fontsize{8}{12}\selectfont
  \caption{Performance Comparison of Different Structures under Different Sampling Rates.}
  \label{performance_comparison}
  \scalebox{1.2}{
    \begin{tabular}{|c|c|c|c|}
    \hline
    \multirow{3}{*}{\tabincell{c}{Sampling \\ Rate}}&
    \multicolumn{3}{c|}{Method}\cr\cline{2-4}
    & \tabincell{c}{ODE-RK3\\(Loss / NMSE)} &\tabincell{c}{ODE-LF\\(Loss / NMSE)}&\tabincell{c}{CNN\\(Loss / NMSE)}\cr
    \hline
    \hline
        {1/2}&{\bf 0.00001 / -39.59dB}  &   0.00002 / -35.53dB    &   0.00003 / -32.37dB\cr\hline
        {1/4}&{\bf 0.00002 / -33.77dB} &   0.00009 / -28.18dB &   0.00012 / -26.78dB\cr\hline
        {1/8}&{\bf 0.00086 / -18.12dB}  &   0.00133 / -16.25dB &   0.00151 / -15.69dB\cr\hline
        {1/16}&{\bf 0.0155 / -5.7dB}   &   0.01834 / -4.95dB    &   0.01929 / -4.69dB\cr
    \hline
    \end{tabular}}
\end{table}

\section{{Simulation Results}}
\begin{figure}[!t]
	\centering
	\includegraphics[width= 100 mm]{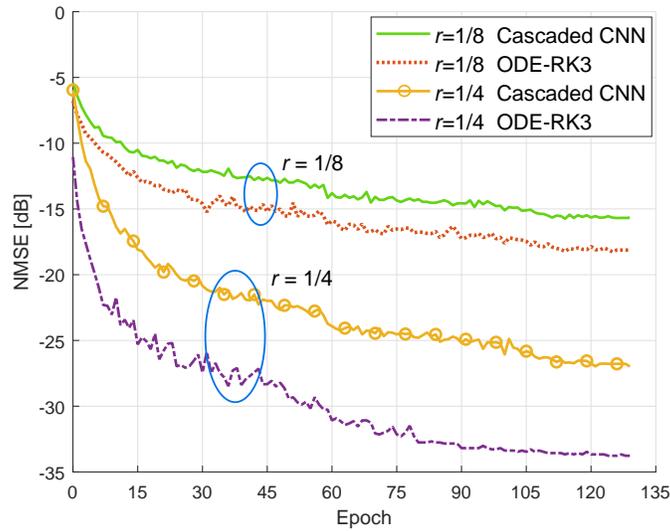}
	\caption{{The NMSE of channel extrapolation versus epochs.}}
	\label{epoch_nmse}
\end{figure}
In this section, we evaluate the channel extrapolation performance of ODE-based CNN through numerical simulation.
We first describe the communication scenario and dataset source, and then show the parameters of the training network. Finally, we show the simulation results and explain the performance of our proposed network.

The scenario we consider is an indoor scene with user, BS and RIS. To generate this scenario, we resort to the indoor distributed massive MIMO scenario ¡®I1¡¯ of the DeepMIMO dataset, which is
generated based on the Wireless InSite \cite{DeepMIMO}.

The ULA at BS has $4$ antennas, i.e., $M = 4$, while the size of the RIS's UPA is $8\times 8$, i.e., $L = 64$.
The carrier frequency of channel estimation is $2.4$ GHz. The OFDM signal bandwidth is set as $20$ MHz,
while the number of subcarriers is $K=64$.
The antenna spacing is $\frac{\lambda}{2}$, and the number of paths is $5$.
Furthermore, the activated users are located from the $1$-st row to the $100$-th row.
Each row contains $201$ users, and the total number of users is $20100$.
The users are split in two parts, i.e., the training and the test groups, according to the ratio $4 : 1$. The sampling rate $r=\frac{N}{L}$ is separately set as $\frac{1}{2}, \frac{1}{4}, \frac{1}{8}$ and $\frac{1}{16}$.


In the simulations, we adopt three network structures for comparison, i.e., the ODE-RK3 structure formed by some RK3-Blocks, the ODE-LF structure containing several LF-Blocks and the cascaded CNN network. For fairness, all CNNs have $26$ layers and the same number of parameters. The ODE-based network contains $4$ RK3-Blocks or LF-Blocks (each block consists of $6$ convolutional layers), a head convolutional layer and a tail convolutional layer.
Considering ODE-RK3 structure as an example, we list the layer parameters of the CNN in TABLE \ref{layer parameters}.
Specially, in the hidden layers, the number of neurons is  $128$, and $\mathrm{ReLU}$ is adopted as the activation function, i.e., $\mathrm{ReLU}(x) = \max(x, 0)$. The kernel size of the first convolutional layer is $5\times 5$, and that of the remainder convolutional layers is set as $3\times 3$. The  learning rate $\eta$ is initialized as $0.0005$ and decreases with increased iteration times. Specifically, after $40$ iterations, the learning rate reduces by $20\%$ for every $10$ epochs.

TABLE \ref{performance_comparison} shows the performance of different network structures.
It can be noted that the proposed ODE CNN is always superior to
the cascaded CNN network, and this gain enhances with the increase of the sampling rate $r$.
The RK3 structure with three-order performs better than the third-order LF structure.
 When sampling rates are $\frac{1}{2}$ and $\frac{1}{4}$, the ODE-RK3 structure can achieve satisfactory results.
 Furthermore, in terms of channel extrapolation normalized MSE (NMSE), the ODE-RK3 network at $r = \frac{1}{4}$ performs better than the CNN network at $r = \frac{1}{2}$, which means that the length of the pilot for the sub-sampled channel $\widetilde {\mathbf C}$ estimation can be reduced through introducing the ODE structure.
 If the compression ratio is relatively low, such as $r = \frac{1}{16}$, the performance of the ODE-based CNN is not significantly better than that of the cascaded CNN due to the reduced raw input information for the channel extrapolation.

\begin{figure}[!t]
	\centering
	\includegraphics[width= 100mm]{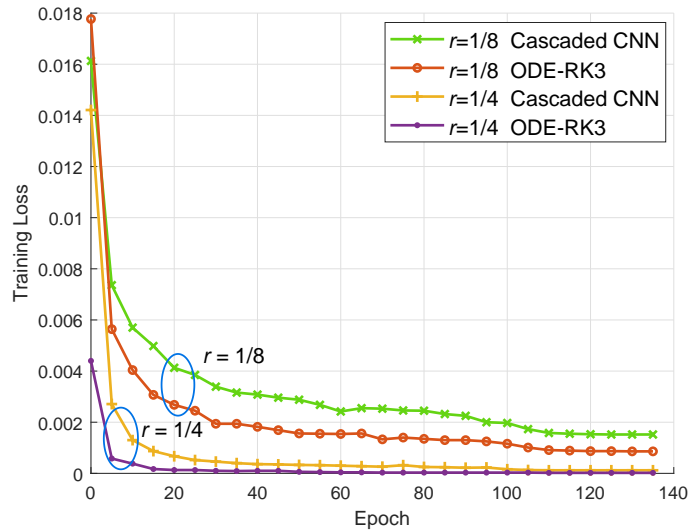}
	\caption{{The loss of channel extrapolation versus epochs.}}
	\label{epoch_loss}
\end{figure}
Fig. \ref{epoch_nmse} depicts the curves of NMSE with respect to the number of epochs.
Two structures (the ODE-RK3 and cascaded CNN structures)
and two sample rates ($\frac{1}{4}$ and $\frac{1}{8}$) are considered here.
It can be seen that with the increase of iteration time, all the NMSE curves present a downward trend and reach the stable levels after $115$ epochs. Furthermore, for a given sampling rate, the NMSEs of the ODE-based CNN are always lower than those of the cascaded CNN.
Fig. \ref{epoch_loss} depicts the training loss of different CNN structures versus epochs. It can be checked from Fig. \ref{epoch_loss} that, with fixed rate $r$, the training loss in the ODE-RK3 network decreases faster than that in the cascaded CNN network, which means that the ODE-based network can be trained
more quickly than CNN.

\begin{figure}[!t]
	\centering
	\includegraphics[width= 100mm]{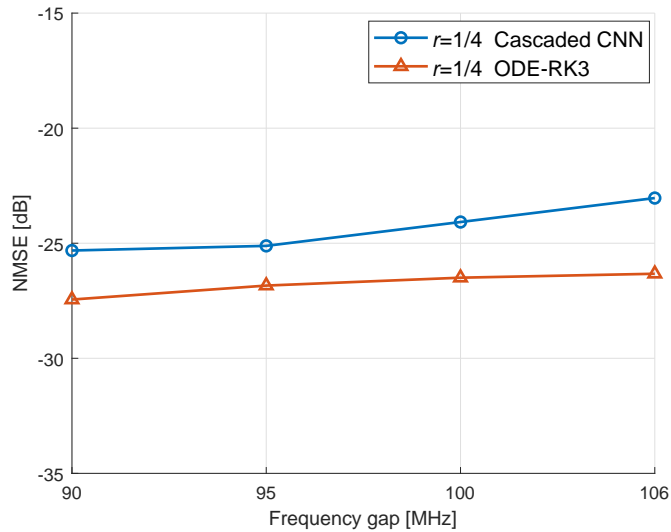}
	\caption{{The NMSEs of the channel extrapolation verus the frequency gaps.}}
	\label{frequency_gap}
\end{figure}
So far, we considered the channel extrapolation at the same frequency band.
However, our proposed scheme can still be used for the case with frequency difference. In actual systems, such as the frequency division duplexing system, the uplink and downlink channels operate in different frequency bands.
Fig. \ref{frequency_gap} shows the performance of the ODE-RK3 and the cascaded CNN
structures under different frequency gaps.
As can be seen from Fig. \ref{frequency_gap}, both ODE-RK3 and CNN structures are affected by the frequency gaps. As the frequency difference increases, the NMSE of channel extrapolation slightly augments.
It is worth noting that the ODE-RK3 structure always performs better than cascaded CNN, which proves the stability and effectiveness of the proposed ODE-based CNN.

\section{Conclusion}
In this paper, we have examined a RIS-assisted MIMO communication system, and designed an ODE-based CNN to extrapolate the cascaded channel. In our scheme, only part of the full CSI is needed. Hence, some of the RIS elements could be turned off through spatial sampling, which greatly reduces the length of the pilot sequence in the channel estimation phase and improves the resource utilization.
Simulation results have demonstrated that
 the proposed extrapolation scheme can effectively compress the large-scale RIS channel over the physical space.
 Moreover, the ODE-based structure can speed up the convergence and improve the performance of the cascaded CNN.
\vspace{-3mm}


\balance

\end{document}